\documentclass[12pt]{article}

\usepackage{epsf}

\def\nn{\nonumber\\}
\def\calD{{\cal D}}

\def\myfrac#1#2{\hbox{\large ${#1\over #2}$}}

\def\JT{{\cal T}}

\def\M{\Sigma }
\def\h{{\rm h}}
\def\t{{\rm t}}
\def\calM{{\cal M}}

\makeatletter
\@addtoreset{equation}{section}

\renewcommand{\thefootnote}{\fnsymbol{footnote}}
\makeatother

\begin{document}
\thispagestyle{empty}

\begin{flushright}
TIT/HEP--511 \\
{\tt hep-th/0310130} \\
October, 2003 \\
\end{flushright}
\vspace{3mm}

\begin{center}
{\Large
{\bf 
Massless Localized Vector Field 
 on a Wall 
} 
\\
\vspace{2mm}{\bf 
in $D=5$ SQED with Tensor Multiplets 
}
} 
\\[12mm]
\vspace{5mm}

\normalsize
  {\large \bf 
Youichi~Isozumi}
\footnote{\it  e-mail address: 
isozumi@th.phys.titech.ac.jp
}, 
  {\large \bf 
 Keisuke~Ohashi 
}\footnote{\it  e-mail address: 
keisuke@th.phys.titech.ac.jp
}, 
~and~~  {\large \bf 
Norisuke~Sakai}
\footnote{\it  e-mail address: 
nsakai@th.phys.titech.ac.jp
} 

\vskip 1.5em

{ \it Department of Physics, Tokyo Institute of 
Technology \\
Tokyo 152-8551, JAPAN  
 }
\vspace{15mm}
{\bf Abstract}\\[5mm]
{\parbox{13cm}{\hspace{5mm}
Massless localized vector field is obtained in 
five-dimensional 
supersymmetric (SUSY) QED coupled to tensor multiplets 
as a half BPS solution. 
The four-dimensional gauge coupling is obtained as a 
topological charge. 
We also find all the (bosonic) massive modes exactly 
for a particular value of a parameter, demonstrating 
explicitly the existence of a mass gap. 
The four-dimensional Coulomb law is shown to hold for sources 
placed on the wall. 
}}
\end{center}
\vfill
\newpage
\setcounter{page}{1}
\setcounter{footnote}{0}
\renewcommand{\thefootnote}{\arabic{footnote}}

\section{Introduction}\label{INTRO}

Brane-world scenario has raised a new possibility to obtain 
unified theories beyond the standard model 
\cite{HoravaWitten, LED, RS}. 
To realize the brane-world scenario, it is necessary to localize 
standard model particles on topological defects such 
as domain walls. 
It has been a long-standing problem to obtain a massless 
vector field localized on a wall. 
If we implement the Higgs mechanism in the bulk and 
restore the gauge symmetry on the wall, we can in fact 
localize the vector field on the wall. 
However, it has been pointed out that superconducting 
bulk will absorb flux coming out of the source 
placed on the wall. 
Therefore these flux will not reach beyond the width of the 
wall even in the direction along the world volume of the wall. 
This screening implies that the 
vector field should have a mass of the order of 
the inverse width of the wall \cite{DvaliShifman, LED}. 
This general argument are confirmed by 
explicit examples such as \cite{MaruSakai, IOS1}. 
Then one is naturally led to consider a dual picture as 
an appropriate setting. 
If a vector field is confined in the bulk 
and deconfined on the wall, the flux coming out of a source 
should be repelled from the wall, producing a four-dimensional 
Coulomb law in the world volume of the wall. 
This mechanism of massless localized vector field has been 
argued, and a toy model in four-dimensions has been proposed 
 \cite{DvaliShifman}. 
This is a nice general idea, but a concrete model 
had to use the nonperturbative effect 
to confine vector field, which is not at all obvious to work 
in higher dimensions such as five dimensions. 
Another mechanism that has been proposed was to use gravity. 
It has been shown that vortex together with the warp 
factor of gravity are needed to obtain a massless 
localized vector field \cite{Oda, DRT, RandjbarShaposhnikov}. 
It is perhaps more desirable to obtain a model which gives 
a massless localized vector field even in the limit of vanishing 
gravitational coupling, since 
the gravitational effects are known to be small. 
Another model \cite{Akhmedov} generalized the idea of 
induced gauge field by quantum effects \cite{DGS}. 
This idea is old \cite{Akama} and attractive, but is rather 
difficult to obtain a reliable approximation scheme for such a 
quantum effects. 

Taking a massive ${\cal N}=2$ 
SUSY QED in four-dimensions as a toy model, 
it has been argued that a 
massless localized gauge field $W_\mu $ is obtained 
by dualizing the massless Nambu-Goldstone scalar 
$\phi$ in the three-dimensional effective theory 
$2\partial_{[\mu } W_{\nu ]} 
= \epsilon_{\mu \nu \rho} \partial^\rho \phi$ 
\cite{ShifmanYung}
\footnote{
Antisymmetrization of indices are denoted by brackets. 
We use the convention to divide by the number of terms in the 
antisymmetrization such as $\partial_{[M } W_{N ]} \equiv 
(\partial_{M } W_{N }-\partial_{N } W_{M })/2$.
}. 
This is certainly an intriguing result, but is difficult 
to generalize to our realistic situation of higher dimensions, 
since the gauge field is dual of 
a compact scalar only in three-dimensional space-time. 
In five dimensions\footnote{
We will denote the five-dimensional indices by capital Latin characters 
$M , N =0, 1, \cdots, 4$ and four-dimensional indices by 
Greek characters $\mu, \nu=0, 1, 2, 3$. 
}, a straightforward application of 
the electromagnetic duality for a vector field 
$W_M $ should give a tensor 
(two form) field $B_{M N }$ 
\begin{equation}
F_{M  N  L }(B) 
= {1 \over 2}\epsilon_{M N L P Q} F^{P Q }(W), 
\label{eq:tensor-vector-dual}
\end{equation}
where the field strengths are defined by 
\begin{equation}
F_{M  N  L }(B)\equiv 3\partial _{[M  }B_{N  L ]}, 
\quad 
F_{M  N   }(W)\equiv 2\partial _{[M  }W_{N  ]}. 
\label{eq:field-strength}
\end{equation}
An interesting 
approach has been proposed using tensor field to obtain a 
massless localized 
vector field \cite{DubovskyRubakov}. 
They assumed 
that a tensor field in five dimensions couples to 
some physically motivated wall configuration 
which is given and fixed externally, and 
argued for a (quasi-)localization and the charge 
universality in 
generic terms. 

The purpose of our paper is to give a concrete 
supersymmetry (SUSY) model 
(with eight supercharges) 
including tensor multiplets together with a vector and 
hypermultiplets in five dimensions, 
in order to construct a fully consistent model 
for a massless localized vector field. 
We find a wall and a massless vector multiplet 
localized on the wall as a consistent solution of 
the equations of motion. 
Since our wall configuration preserves half of SUSY 
\cite{IOS1}, we obtain a Massless $U(1)$ vector multiplet 
in the ${\cal N}=1$ SUSY four-dimensional effective theory. 
We find that the four-dimensional gauge coupling is expressed 
as a topological charge associated with the wall. 
Moreover, we obtain not only the massless mode but also 
massive modes of the vector multiplet exactly in one choice of 
a parameter. 
By introducing a static source, we show that the 
four-dimensional 
Coulomb law of the usual minimal electromagnetic interaction 
is reproduced. 
Our mechanism has some similarities to that in 
Ref.\cite{DubovskyRubakov}, such as the 
generic nature of the 
mechanism of the massless localized vector multiplet. 
However, we have a fully consistent model including 
the wall and the massless localized vector field as 
solutions of equations of motion, without an ad hoc 
assumption for the wall as a given external configuration. 
Moreover we start from a SUSY theory in five dimensions, 
resulting in an ${\cal N}=1$ SUSY effective low-energy 
theory.

In Sect.\ref{sc:model}, our model with the tensor multiplet 
is introduced. 
In Sect.\ref{sc:massless-vector}, massless localized vector 
field is obtained. All the massive modes are also 
found for a particular value of a parameter. 
An effective Lagrangian containing all massive modes 
is also worked out to the quadratic order. 
In Sect.\ref{sc:coulomb}, the four-dimensional Coulomb 
law is obtained between sources placed on the wall. 
In Sect.\ref{sc:discussion}, a possible generalization 
(without SUSY) to arbitrary 
space-time dimensions is proposed and a number of 
remaining issues are noted. 
Some details of massive mode functions are given in Appendix. 

\section{Our Model with Tensor Multiplets}
\label{sc:model}

It has been known that tensor multiplets 
can couple to vector multiplets in five-dimensional 
SUSY theories. 
On the other hand, 
a five-dimensional 
SUSY model with hypermultiplets coupled to 
a $U(1)$ vector multiplet can give a domain wall 
as a half BPS solution, 
producing a wall configuration for the vector multiplet 
scalar $\Sigma$ \cite{IOS1}. 
To build a wall, we introduce 
a $U(1)$ vector multiplet, whose bosonic components are 
gauge field $W_M $, scalar field $\Sigma$, and 
$SU(2)_R$ triplet of auxiliary fields $Y^{a}, a=1,2,3$. 
We also need hypermultiplets, whose bosonic 
components are $SU(2)$ doublets of scalar fields $H^{iA}$, 
and auxiliary fields $F_i^{A}$, with the $i=1, 2$ and $A$ 
is the flavor indices of hypermultiplets. 
The number of SUSY vacua is equal or less than 
the number of hypermultiplets. 
To obtain a single wall solution, we take the number of 
hypermultiplets to be two, for simplicity : $A=1, 2$. 
It has been shown that these two types of multiplets 
suffice to produce a wall in five dimensions 
preserving half of SUSY (a $1/2$ BPS state) \cite{IOS1}. 
Our Lagrangian consists of two terms, ${\cal L}_{\rm wall}$ 
to produce a wall, and ${\cal L}_{\rm T}$ to obtain 
a coupling of tensor multiplets with the vector multiplet 
\begin{eqnarray}
 {\cal L}_{\rm total}={\cal L}_{\rm wall}+{\cal L}_{\rm T} \,. 
\label{ToTaction}
\end{eqnarray}

As a concrete example, bosonic part of our Lagrangian 
for the wall reads 
\begin{eqnarray}
\left.{\cal L}_{\rm wall}\right|_{\rm bosonic}
&\!\!\!=&\!\!\!-{1 \over 4}F_{M N }(W) F^{M N }(W)+ 
{1 \over 2}\partial^M  \Sigma \partial_M  \Sigma 
+
{\cal D}_M  H_{iA}^\dagger {\cal D}^M  H^{iA} 
- H_{iA}^\dagger (g_{\rm h} \Sigma -m_A)^2 H^{iA} 
\nonumber\\
&\!\!\!&\!\!\!+\frac{1}{2}(Y^a)^2 - g_{\rm h} \zeta^aY^a 
+ H^\dagger_{iA}(\sigma^a g_{\rm h} Y^a)^i_jH^{jA} +F^{\dagger i}_AF_i^A ,
\label{eq:wall-lagrangian}
\end{eqnarray}
where $g_{\rm h}$ denotes the hypermultiplet gauge coupling 
including its charge, ${\cal D}_M =\partial_M  +ig_{\rm h} W_M $, 
and covariant derivative and $\zeta^a$ are the $SU(2)_R$ 
triplet of Fayet-Iliopoulos parameters\footnote{
We changed the normalization of the vector multiplet by $g_h$ 
from Ref.\cite{IOS1} to make the kinetic term canonical. 
}. 
Without loss of generality, we assume $\zeta^a=(0,0,\zeta>0)$, 
and $m_A=(m_{\rm h}, -m_{\rm h})$. 
It has been known for sometime that the above Lagrangian 
admits BPS single and multiple domain wall solutions 
in the limit of 
infinite gauge coupling, where the vector multiplet becomes 
just a Lagrange multiplier and the model reduces to a 
nonlinear sigma model with only hypermultiplets as physical 
degrees of freedom 
\cite{GTT, EFNS}. 
However, we wish to retain the vector multiplet as a dynamical 
degree of freedom, rather than a Lagrange multiplier field. 
For that purpose, recently obtained exact solutions 
of BPS domain walls for discrete finite values of 
gauge coupling are extremely useful \cite{IOS1}. 
As the simplest case, 
we have shown that 
the exact solution of 
 a single wall 
is obtained for a finite coupling 
\begin{equation}
g_{\rm h}^2 \zeta = 2 m_{\rm h}^2.
\end{equation} 
Apart from the vicinity of the wall, the charged 
hypermultiplets 
takes nonvanishing value, and the $U(1)$ gauge symmetry 
is broken 
spontaneously in the bulk. 
The vector multiplet scalar $\Sigma$ also exhibits 
a kink-like behavior interpolating between two vacua 
$\Sigma = \pm {m_\h\over g_\h}$ 
\begin{eqnarray}
 \M={m_\h\over g_\h}\tanh(m_\h y).  
\label{Mconfig}
\end{eqnarray}
We find that the energy density is concentrated around $y=0$ 
\cite{IOS1}. 

We wish to emphasize the general nature of our mechanism 
to localize a massless vector field. 
We only need a wall configuration for the scalar field $\Sigma$ 
of a vector multiplet which couples to our tensor multiplets. 
Our explicit model for a wall is just to show that there is a 
consistent theory including all ingredients, in particular 
with SUSY. 
To emphasize this point, 
in the most part of this paper, 
we use only the following information on the background 
domain wall configuration : 
background value of the scalar field of the vector multiplet, 
$\langle \M\rangle $, satisfies the BPS equation, 
\begin{eqnarray}
  \langle \M\rangle '\equiv {d \langle \M\rangle \over d y}
=\langle Y^3\rangle ,\quad \langle Y^1\rangle 
=\langle Y^2\rangle =0, \label{eq:BGV}
\end{eqnarray}
where\footnote{
Our convention is 
$\epsilon_{12}=-\epsilon_{21}=\epsilon^{12}=-\epsilon^{21}=1$. 
The $SU(2)_R$ indices are raised and lowered by contracting 
upper left with lower right indices as 
$Y^{ij}=\epsilon^{jk}Y^i{}_k=\epsilon^{ik}Y_k{}^j$. 
}, $Y^{ij}=\sum_{a=1}^3Y^a(i\sigma ^a)^i{}_k\epsilon ^{jk}$, 
and the $\langle \M\rangle $ depends on only the coordinate $y$, 
and approaches two different values at 
left and right spacial infinity $y\rightarrow \pm \infty$, 
like the configuration (\ref{Mconfig}). 
Furthermore, we assume the four-dimensional Lorentz 
invariance on the world volume, resulting in 
$\langle W_M  \rangle 
=0$.
Whenever an explicit model becomes useful, we will always use our 
simplest exact solution in Eq.(\ref{Mconfig}).

In an off-shell formulation of SUSY (and Supergravity) in five 
dimensions, two supermultiplet containing the antisymmetric 
tensor field $B_{M  N  }$ can appear.  
One of the supermultiplets is called tensor gauge multiplet, 
whose $B_{M  N  }$ is massless and admits 
gauge transformations by one-form, 
$\delta B_{M  N  }=2\partial _{[M  }\Lambda _{N  ]}$. 
The other supermultiplet is called 
the large (massive) 
tensor multiplet, 
whose 
bosonic components are antisymmetric tensor (two-form) 
field $B_{M N }^\alpha$, scalar field $\sigma^\alpha$, 
and $SU(2)_R$ triplet of auxiliary fields $X^\alpha_{ij}=X^\alpha_{ji}$, 
where $i, j=1, 2$. 
They have to come in pairs, but we will see 
that one of the two can be interpreted as an auxiliary field.
A pair of the large tensor multiplet 
$T^\alpha =
(\sigma ^\alpha ,\,B_{M  N  }^\alpha ,\,X^{\alpha ij}
), \, (\alpha =1,2)$ 
can have a mass term, and can carry a 
charge $g_{\rm t}$ for the $U(1)$ gauge field $W_M $.   
In our model, we use a pair of the large tensor multiplet as 
a minimal model. 
By now it is well-established that the most general Lagrangian 
for tensor and vector multiplets in five dimensions is characterized 
by a nonlinear kinetic term which is specified by 
second derivatives of a norm function ${\cal N}$ 
which is at most cubic in tensor and vector multiplets 
\cite{Seiberg, ref:Tensor, KugoOhashi}. 
The invariance under the $U(1)$-gauge transformation 
with a gauge parameter $\Lambda $ 
for this multiplet  
\begin{eqnarray}
 \delta (\Lambda )T^\alpha 
=-\Lambda g_\t\epsilon _{\alpha \beta }T^\beta ,
\label{Gtrf} 
\end{eqnarray}
determines this cubic term of the norm function. 
This allows the tensor multiplets to interact with the vector 
multiplet, which carries informations 
of the domain wall configuration. 
By defining 
\begin{equation}
\calM\equiv m_\t-g_\t\M, 
\end{equation}
we obtain the bosonic part of our Lagrangian 
containing tensor multiplets ${\cal L}_{\rm T}$ as 
\begin{eqnarray}
\left.{\cal L}_{\rm T}\right|_{\rm bosonic}
&\!\!\!=&\!\!\!\calM \sum_{\alpha =1}^2\left(
-\myfrac14 B_{M  N  }^\alpha  B^{M  N  \alpha }
+\myfrac12\calD^M  \sigma ^\alpha \calD_M  \sigma ^\alpha 
+\myfrac14X^{ij \alpha }X^{\alpha }_{ij}
-\myfrac12\calM^2(\sigma ^\alpha )^2\right)\nn
&\!\!\!&\!\!\!{}-\sum_{\alpha =1}^22g_\t\sigma ^\alpha 
\left(\myfrac14 B_{M  N  }^\alpha  F^{M  N  }(W)
+\myfrac12\calD^M  \sigma ^\alpha \partial _M  \M
+\myfrac14X_{ij}^\alpha Y^{ij}\right)\nn
&\!\!\!&\!\!\!{}-\myfrac18\epsilon ^{M  N  L P Q }
B_{M  N  }^\alpha \partial _L 
B_{P Q }^\beta \epsilon _{\alpha \beta }
-\myfrac18g_\t\epsilon ^{M  N  L P Q }
W_L B_{M  N  }^\alpha B_{P Q }^\alpha,  \nn
\label{eq:Taction}
\end{eqnarray}
where, our convention of the space-time metric is 
$\eta _{M  N  }={\rm diag}(1,-1,-1,-1,-1)$, and we omitted the auxiliary
fields\footnote{
They are obtained as transformations by the central charge 
$Z\sigma ^\alpha ,Z^2\sigma ^\alpha $ \cite{KugoOhashi}. 
}, which are not important for our model.
The mass parameters $m_\t$ and the charge $g_\t$ are arbitrary 
at this point. 
However, we will later find that one of 
these two parameters must be tuned to assure 
the expected mechanism to work.
The covariant derivative $\calD_M  \sigma ^\alpha $ 
of the scalar field $\sigma^\alpha$ 
of the tensor multiplet 
is defined as usual  
\begin{eqnarray}
 \calD_M  \sigma ^\alpha 
=\partial _M  \sigma ^\alpha 
-g_\t W_M  \epsilon _{\alpha \beta }\sigma ^\beta .
\end{eqnarray}
The five-dimensional Lagrangian 
${\cal L}_{\rm total}={\cal L}_{\rm wall}+{\cal L}_{\rm T}$ 
given in (\ref{eq:wall-lagrangian}) and (\ref{eq:Taction}) 
together with its fermionic terms are invariant 
under the five-dimensional SUSY transformation 
(with eight supercharges) . 
It contains kinetic terms of the 
2-form tensor fields $B_{M  N  }$, as well as other fields.

Let us rewrite the Lagrangian (\ref{eq:Taction}) by integrating out the
auxiliary field, $X_{ij}^\alpha $ and considering fluctuations around the
vacuum expectation values.
An equation of motion for the $X_{ij}^\alpha $  can be read as 
\begin{eqnarray}
 \calM\, X^\alpha _{ij}-g_\t \sigma ^\alpha Y_{ij}=0.
\end{eqnarray}
The equations of motion of the scalar $\sigma ^\alpha $ are 
\begin{eqnarray}
0&\!\!\!=&\!\!\!
 \calD^M  (\calM\calD_M  \sigma ^\alpha )+\sigma ^\alpha \partial ^M  \partial _M  \calM+\calM^3\sigma ^\alpha 
+\myfrac12g_\t B_{M  N  }^\alpha F_{M  N  }(W)
+\myfrac12g_\t X^\alpha _{ij}Y^{ij}. 
\end{eqnarray}
These equations are consistent with  
\begin{eqnarray}
 \langle \sigma ^\alpha \rangle 
=\langle B_{M  N  }^\alpha \rangle 
=\langle X^\alpha _{ij}\rangle =0, 
\label{eq:BGT}
\end{eqnarray}
which can also be derived from the requirement of 
the BPS condition preserving half of the eight SUSY. 
Namely the BPS wall solution (\ref{Mconfig}) or (\ref{eq:BGV}) 
of the Lagrangian ${\cal L}_{\rm wall}$ is not disturbed by adding 
the tensor multiplet to the system. 

The quadratic terms of the fluctuations of the fields 
around the background (\ref{eq:BGV}), 
(\ref{eq:BGT}) can be read as
\begin{eqnarray}
\left.{\cal L}_{\rm T}\right|_{\rm bosonic}
&\!\!\!=&\!\!\!
\langle \calM\rangle \sum_{\alpha =1}^2\left(
-\myfrac14 B_{M  N  }^\alpha  B^{M  N  \alpha }+\myfrac12 \partial ^M  \sigma ^\alpha \partial _M  \sigma ^\alpha 
\right)\nn
&\!\!\!&\!\!\!{}-\sum_{\alpha =1}^2\myfrac12\left(
\langle \calM\rangle ^3+{(\langle \calM\rangle ')^2 \over \langle \calM\rangle }
-\langle \calM\rangle ''\right)(\sigma ^\alpha )^2\nn
&\!\!\!&\!\!\!
{}-\myfrac18\epsilon ^{M  N L  P Q  }B_{M  N  }^\alpha 
\partial _L B_{P Q  }^\beta 
\epsilon _{\alpha \beta }+\sum_{\alpha =1}^2\left(-\myfrac12
\langle \calM\rangle '(\sigma ^\alpha )^2\right)'
\nn
&\!\!\!&\!\!\!{}+(\hbox{higher order terms of the fluctuations})
\label{eq:Taction2},
\end{eqnarray}
where we used 
$\langle \calM\rangle '
=-g_\t\langle \M\rangle '=-g_\t\langle Y^3\rangle $.
We consider only this quadratic part of the Lagrangian 
in the following.

\section{
Massless Localized Vector Field from the Tensor Field}
\label{sc:massless-vector}

\subsection{A Lagrangian for the $2$-Form Tensor Fields}
\label{sc:2form-tensor}

In this section, we concentrate on the kinetic term 
${\cal L}_{\rm 2form}'$ 
for the 2-form tensor fields 
\begin{eqnarray}
 {\cal L}'_{\rm 2form}
&\!\!\!=&\!\!\!
-\myfrac14MB^{1}_{M  N  }B^{1 M  N  }
-\myfrac14 M B^{2}_{ M  N  } B^{2 M  N  }
-\myfrac1{8}\epsilon ^{L M  N  P Q  }
B^{1}_{M  N  }\partial _L B^{2}_{P Q  }
+\myfrac1{8}\epsilon ^{L M  N  P Q  } 
B^{2}_{M  N  }\partial _L B^{1}_{P Q  } \nn
\end{eqnarray}
where we rewrite 
$M\equiv \langle \calM\rangle(y) $ 
for simplicity.  
This Lagrangian is, so called, a self-dual Lagrangian in
five dimensions.  
 Furthermore, we rewrite the Lagrangian as follows  
\begin{eqnarray}
{\cal L}'_{\rm 2form}&\!\!\!=&\!\!\!
{\cal L}_{\rm 2form} +{\cal L}_\theta , \nn
{\cal L}_{\rm 2form}&\!\!\!
=&\!\!\!-\myfrac14M B^{1}_{M  N  }B^{1 M  N  }
-\myfrac14 M B^{2}_{M  N  } B^{2 M  N  }
+\myfrac1{4}\epsilon ^{L M  N  P Q  }
 B^{2}_{ M  N  }\partial _L B^{1}_{ P Q  },\nn
{\cal L}_\theta &\!\!\!=&\!\!\!\partial _L 
\left(-\myfrac18\epsilon ^{L M  N  P Q  }
B^{1}_{M  N  } B^{2}_{P Q  }\right) . 
\end{eqnarray}
In a space extending to infinity without boundaries, we can 
freely use either one of these Lagrangians ${\cal L}'_{\rm 2form}$ 
and ${\cal L}_{\rm 2form}$, since total divergence term 
${\cal L}_\theta$ does not contribute to the action. 
However, we are considering a wall soution which 
approaches to different vacua 
at left and right infinities, respectively, 
resulting in a topologically nontrivial configuration. 
Moreover, the $U(1)$ gauge invariance is spontaneously 
broken in the bulk. 
Therefore we need to decide  how much total divergence terms 
should be included in our fundamental Lagrangian. 

In order to extract physics out of our model, we note that 
one of the two-form tensor field should be treated as 
an auxiliary field. 
For instance, by varying with respect to $ B^{2}_{M N }$, 
we obtain the equation of motion for $ B^{2}_{M N }$ as 
\begin{eqnarray}
0&\!\!\!=&\!\!\!
-M B^{2 M  N  }
+\myfrac1{2}\epsilon ^{M  N L  P Q  }
\partial _L B^{1}_{P Q  }, 
\label{eq:barB-EOM}
\end{eqnarray} 
which allows to express $ B^{2}_{M N }$ algebraically 
in terms of $B^{1}_{M N }$. 
If we start from the Lagrangian ${\cal L}_{\rm 2form}$, 
we can derive the above equation without performing 
a partial integration. 
If we start from any other Lagrangian, such as ${\cal L}'_{\rm 2form}$ 
instead, we first need to add a total divergence term 
${\cal L}_\theta$ and obtain the Lgrangian 
${\cal L}_{\rm 2form}$, 
so that we can derive the equation of motion (\ref{eq:barB-EOM}).  
In order to interpret one of the tensor field $B^{2}_{M N }$ 
as an auxiliary field, we decide to choose 
${\cal L}_{\rm 2form}$ as our fundamental Lagrangian, 
and denote the remaining tensor field as 
$B_{MN} \equiv B^{1}_{MN}$ from now on. 

If we consider a space-time with boundaries, for instance 
at $y=\pm \pi L$, 
applying the $U(1)$ gauge transformations (\ref{Gtrf}) on 
our fundamental Lagrangian gives 
\begin{eqnarray}             
 \delta (\Lambda )\int ^{\pi L}_{-\pi L}dy
 \int d^4x{\cal L}_{\rm 2form}
&\!\!\!=&\!\!\!
\int d^4x\left[g_\t\Lambda 
\myfrac18\epsilon ^{y\mu\nu \rho\sigma}
(B^{1}_{\mu\nu }B^{1}_{\rho\sigma}- B^{2}_{\mu\nu }
 B^{2}_{\rho\sigma})\right]^{\pi L}_{-\pi L}, 
\end{eqnarray}
which vanishes if  the fields are periodic 
($B^{1}_{M N }(y=-\pi L)=B^{1}_{M N }(y=\pi L),  
 B^{2}_{M N }(y=-\pi L)= B^{2}_{M N }(y=\pi L)$). 
 For topologically nontrivial situations, 
fields are no longer periodic, and 
the $U(1)$ gauge invariance are broken 
on the boundaries \footnote{
This is somewhat reminiscent of the Chern-Simons theory in 
three dimensions \cite{Witten}. 
} : 
the parameter of the transformation on the boundary must vanish 
on the boundaries 
($ \Lambda (\pi L)=\Lambda (-\pi L)=0$). 
It is gauge invariant if there is no boundaries. 
If there is a boundary, the gauge degrees of freedom emerge as 
conformal field theories \cite{EMSS}.

Since the Lagrangian ${\cal L}_{\rm 2form}$ is only quadratic 
in $ B^{2}_{M N }$, we can perform the functional integral 
of $B^{2}_{M N }$ exactly. 
The quadratic part of the resulting Lagrangian is 
now written in terms of $B_{MN} \equiv B^{1}_{MN}$ only 
\begin{eqnarray}
 {\cal L}_{\rm 2form}
&\!\!\!=&\!\!\!{1\over 12 M(y)}F_{M  N L  }(B)F^{M  N L  }(B)
-\myfrac14M(y) B_{M  N  }B^{M  N  }.
\label{eq:quad-tensor-lag}
\end{eqnarray}
This Lagrangian plays the most important role in our paper. 
If the mass function 
$M(y)=\langle {\cal M} \rangle
=\langle m_{\rm t}-g_{\rm t}\Sigma(y) \rangle$ is a constant, 
 this Lagrangian is reduced to an ordinary kinetic term for a massive
 tensor field, where no massless state is contained in the tensor field. 
Dubovsky and Rubakov observed, however, 
if one takes a limit of $M\rightarrow +0$ with 
$B_{M  N  }/\sqrt{M}$ fixed, 
the Lagrangian is reduced to the kinetic term for
 a 2-form gauge tensor field, that is, a massless field 
\cite{DubovskyRubakov}. 
In our case, the scalar field $\Sigma(y)$ gives $M(y)$ 
a non-trivial dependence on $y$, 
which produces a region where $M(y)$ vanishes 
approximately. 
Therefore we can expect that the 
 massless vector mode may exist in our system.     

Now, let us examine this system in detail. 
By varying $B_{M N }$, we obtain the equation of motion of 
the physical field $B_{M  N  }$ as 
\begin{eqnarray}
 0&\!\!\!=&\!\!\!\partial ^L 
\left({1\over M(y)}F_{M  N L  }(B)\right)
+M(y)B_{M  N  }. 
\label{eomB}
\end{eqnarray}
To do partial integration here, 
a boundary condition is needed to eliminate the surface term 
\begin{eqnarray}
 \int d^4x\left[\delta B^{\mu\nu }
\left({1\over M(y)}
F_{\mu\nu y}(B)\right)\right]^{y=\infty }_{y=-\infty }=0 .
\label{condboundary}
\end{eqnarray}
 We will study solutions of the equation (\ref{eomB}) under the condition 
 (\ref{condboundary}) in the next context.

\subsection{Massless Modes and Localization}
\label{sc:massless-mode}

The most interesting and important point is 
the question whether the solution of the 
equation (\ref{eomB}) contains a four-dimensional 
massless vector mode or not. 
Since we assume four-dimensional Lorentz invariance on 
the world volume, it is useful to introduce the momentum 
space $p_\mu,(\mu=0,\cdots,3)$ in four dimensions. 
Let us first study the massless modes  $p^2=0$. 
In this case, it is useful to decompose four-dimensional 
Lorentz vectors in terms of the following basis vectors : 
the longitudinal component $p_\mu$, 
 the scalar component $l_\mu$, 
and 
two transverse polarisation components $\epsilon _\mu^i, (i=1,2)$, 
which are defined by 
$l_\mu l^\mu=0,\, l_\mu p^\mu=1$, 
$\epsilon _\mu^i p^\mu=\epsilon _\mu^i l^\mu=0$, 
and 
$\epsilon _\mu^i \epsilon ^{\mu j}=-\delta^{ij}$. 
By substituting these expansions of the field $B_{M N }$ 
to the equation (\ref{eomB}), we obtain solutions as  
\begin{eqnarray}
B_{\mu\nu }(x,y)|_{\rm massless}
&\!\!\!=&\!\!\!\int {d^4p\over (2\pi i)^4}\delta (p^2)
\left(2p_{[\mu}l_{\nu]}a(p)\phi _1(y)
+2ip_{[\mu}\epsilon _{\nu]}^ib_i(p)\phi _2(y)\right.\nn
&\!\!\!&\!\!\!
\qquad \qquad  \left.{}+2il_{[\mu}\epsilon _{\nu]}^ic_i(p)\phi _3(y)
+\epsilon _{[\mu}^1\epsilon _{\nu]}^2d(p)\phi _4(y)\right)
e^{ip_\lambda x^\lambda}\nn
&\!\!\!&\!\!\! {}
+\rho (y)\int {d^4p\over (2\pi i)^4}
\delta (p^2)2ip_{[\mu}\epsilon _{\nu]}^ic_i(p)e^{ip_\lambda x^\lambda},
\label{eq:Bmn-massless}
\\
B_{\mu y}(x,y)|_{\rm massless}
&\!\!\!=&\!\!\!{1\over M(y)^2}\int {d^4p\over (2\pi i)^4}
\delta (p^2)\left(ip_{\mu}a(p)\phi _1'(y)
+\epsilon _{\mu}^ic_i(p)\phi _3'(y)\right)e^{ip_\lambda x^\lambda },
\label{eq:Bmy-massless}
\end{eqnarray} 
where, the four independent mode functions 
$\phi _a(y),\,(a=1,\cdots,4)$ have to satisfy the same equation 
\begin{eqnarray}
 0=\left({\phi '(y)\over M(y)}\right)'-M(y)\phi (y), 
\label{eq:phi-eq}
\end{eqnarray}
whereas another mode function $\rho (y)$ 
must be determined in terms of 
the mode function 
$\phi_3(y)$ as 
\begin{eqnarray}
0=\left({\rho '(y)\over M(y)}\right)'-M(y)\rho (y)
+\left({1\over M(y)^2}\right)'{\phi _3'(y)\over M(y)}.
\end{eqnarray}
If these mode functions are suitably normalizable, the corresponding 
four-dimensional fields $a(p), b_i(p), c_i(p), d(p)$ are physical 
massless fields. 

If $\phi_3(y)$ happens to vanish, the function $\rho(y)$ satisfies 
the same equation as $\phi(y)$. 
Then there is no distinction between the field $c_i(p)$ and the field 
$b_i(p)$, and the term with $\rho(y)$ can be absorbed 
into the term with $b_i(p)$. 
Therefore we can define $\rho(y)=0$ when $\phi_3(y)=0$, and then the field 
$c_i(p)$ does not exist in that case.

The most general solution of Eq.(\ref{eq:phi-eq}) reads 
\begin{eqnarray}
 \phi (y)=C_1e^{s(y)}+C_2e^{-s(y)},\quad s(y)\equiv 
\int ^\infty _{y}dzM(y).
\end{eqnarray}
The behavior of this solution at $y\rightarrow \pm \infty$ 
shows that the scalar field configuration $M(y)$ 
must vanish at $y=\infty $ or $y=-\infty $. 
Otherwise, $\phi (y)$ (that is $B_{M  N  }$) diverges 
at $y=\infty $ or $y=-\infty $. 
Therefore, without loss of generality, we assume that 
\begin{eqnarray}
 M(y)&\!\!\!\rightarrow &\!\!\!
+0, \quad {\rm as }\quad y\rightarrow \infty ,
\label{eq:tunig}
\\
\phi (y)&\!\!\!=&\!\!\!g_e^{-1}e^{-s(y)},\quad (\phi '=M\phi ).
\label{yfunction}
\end{eqnarray}  
where we denote the integration constant as $g_e$, since 
it will play a role of a coupling constant later. 
Note that the mode function $\phi (y)$ approaches 
the value $g_e^{-1}$ asymptotically 
in the region where the background 
$M(y)
=\langle m_{\rm t}-g_{\rm t}\Sigma(y) \rangle$ 
tends to vanish, whereas it vanishes at the opposite infinity. 
On the other hand, we need not solve the equation for the 
$\rho (y)$, as we will find later that it does not provide 
physical normalizable mode.

To find out physical modes with normalizable wave 
functions, we will demand that 
the energy density $T_{00}$ of the system to be bounded 
from above, since we will eventually consider continuum 
massive states as well. 
It is easiest to introduce spacetime metric 
$g_{M N }$ temporarily and to take the flat space limit 
after varying the Lagrangian with respect to $g_{M N }$ 
\begin{eqnarray}
T_{00}\equiv 
\left. 2{\delta S\over \delta g^{00}}\right|_{g^{M  N  }
\rightarrow \eta ^{M  N  }}
=\left.\left[{1\over 12M}F_{M  N L  }(B)^2
+{M\over 4}B_{M  N  }^2\right]\right|_{\eta ^{M  N  }
\rightarrow \delta ^{M  N  }}.
\label{eq:energy-density}
\end{eqnarray}
The result of this manipulation is that the kinetic 
energy density of gauge fields (the first term of the 
right-hand side of Eq.(\ref{eq:energy-density})) 
is given by the sum of 
the square of the time derivatives 
$F_{0\mu\nu}$ (electric field) and 
the square of the spacial derivatives 
$F_{\mu\nu\lambda}$ (magnetic field) 
instead of their difference as in the Lagrangian. 
This point is represented by replacing $\eta^{M N }$ by 
$\delta^{M N }$. 
Therefore both the first and the second term of 
the right-hand side of this 
 equation is positive definite. 
This formula contains the following term 
\begin{eqnarray}
{M\over 2}B_{\mu y}^2|_{\eta \rightarrow \delta } 
\end{eqnarray}
and the contribution from the massless modes to this term 
has the following $y$-dependence 
\begin{eqnarray}
 {M\over 2}\left({\phi '\over M^2}\right)^2={\phi ^2\over 2M},
\end{eqnarray}  
where we used $\phi '=M\phi $, because of the solution 
(\ref{yfunction}). 
Clearly this term diverges at $y\rightarrow \infty $, 
showing the nonnormalizability of the mode. 
Hence, from the requirement of the bounded energy 
density, $c_i(p)$ and $a(p)$ components 
of the massless modes $B_{\mu y}$ in 
Eq.(\ref{eq:Bmy-massless}) 
should not exist as physical fields. 
Therefore, the part $B_{\mu y}$ of tensor field has no massless 
modes. 
Similarly, the following term contained in the energy density 
\begin{eqnarray}
 {1\over 12M}F_{\mu\nu\lambda}(B)^2|_{\eta \rightarrow \delta }
\end{eqnarray}
has the same $y$-dependent contribution from the 
$d(p)$ component in Eq.(\ref{eq:Bmn-massless}). 
Therefore $F_{\mu\nu\lambda}(B)$ has no massless modes, 
that is, the $d(p)$ component field should vanish. 

Therefore, up to this point, only the mode $b_i(p)$ remains as a 
candidate of normalizable massless mode of the system. 
This mode corresponds to the four-dimensional massless vector field 
$A_\mu(x)$ which we anticipated. 
Eq.(\ref{eq:Bmn-massless}) implies that 
the contribution of the component field $b_i(p)$ to the 
tensor field $B_{\mu\nu }$ can be expressed in terms of 
the field strength $F_{\mu\nu }(A)=2\partial _{[\mu}A_{\nu]}$ 
of a vector potential $A_\mu$ and the function
$\phi (y)$ as 
\begin{eqnarray}
 B_{\mu\nu }(x,y)|_{\rm massless}
&\!\!\!=&\!\!\!
\phi (y)F_{\mu\nu }(A(x)),\quad \partial ^\nu F_{\mu\nu }(A)=0,
\label{eq:massless-comp}
\\ 
B_{\mu y}(x,y)|_{\rm massless}&\!\!\!=&\!\!\!0.
\end{eqnarray}
Let us verify that our massless filed candidate $b_i(p)$ really 
gives a bounded energy density. 
The energy density (\ref{eq:energy-density}) can now be given 
by a sum of 
the electric field
$(\vec E)_i=F_{0i}(A)$ and the magnetic field 
$(\vec B)_i=\tilde F_{0i}(A)$ as 
\begin{eqnarray}
T_{00}(x, y)&\!\!\!=&\!\!\!
M(y)\phi ^2(y)\left(\vec E^2(x)+\vec B^2(x)\right)
\equiv {f(y)\over 2{g_e}^2}
\left(\vec E^2(x)+\vec B^2(x)\right),
\label{eq:energy-density-massless}
\end{eqnarray}
where the profile $f(y)$ of the energy density is 
defined as 
\begin{equation}
 f(y)\equiv 2g_e^2M(y)\phi (y)^2.
\label{eq:energy-density-profile}
\end{equation}
The effective 
four-dimensional gauge coupling 
$g_e$ is defined by requiring 
the effective four-dimensional energy which is given by 
integrated over $y$ to be 
\begin{equation}
\int dy T_{00}(x, y) 
= {1 \over 2g_e^2}\left(\vec E^2(x)+\vec B^2(x)\right) . 
\end{equation}
The above result (\ref{eq:energy-density-massless}) 
shows that this effective 
four-dimensional gauge coupling $g_e$ 
is a topological charge, 
which is determined solely by the boundary condition as 
\begin{equation}
\int ^\infty _{-\infty }dy 
2  M(y)\phi (y)^2 
= \int ^\infty _{-\infty }dy 
{d \over d y} \left(\phi(y)^2 \right) 
= \left[\phi (y)^2\right]^\infty _{-\infty }
={1 \over g_e^2}, 
\end{equation}
where we used the equation (\ref{yfunction}).
Thus the profile function of the energy density 
is normalized as 
\begin{eqnarray}
\int ^\infty _{-\infty }dy 
f(y)
=1. 
\end{eqnarray}
We can easily find that the configuration of $f(y)$ 
vanishes at $y\rightarrow \infty $
because of $M(y)\rightarrow 0$, and also vanishes 
at $y\rightarrow -\infty $ because of 
$\phi (y)\rightarrow 0$, and the region 
corresponding to the wall, 
where the 
configuration of 
$ M(y)\equiv \langle {\cal M} \rangle 
=\langle m_\t-g_\t \Sigma \rangle $ varies, gives 
a finite contribution of $f(y)$. 
This behavior of the energy density illustrate 
that {\it the four-dimensional massless vector field $A_\mu$ is
localized on the wall as the solution of the equations of motion 
for the tensor multiplets}.   
 
\begin{figure}[htb]
\begin{center}
\leavevmode
\epsfxsize=6cm
\epsfysize=4cm
\begin{picture}(100,100)(50,10)
\epsfbox{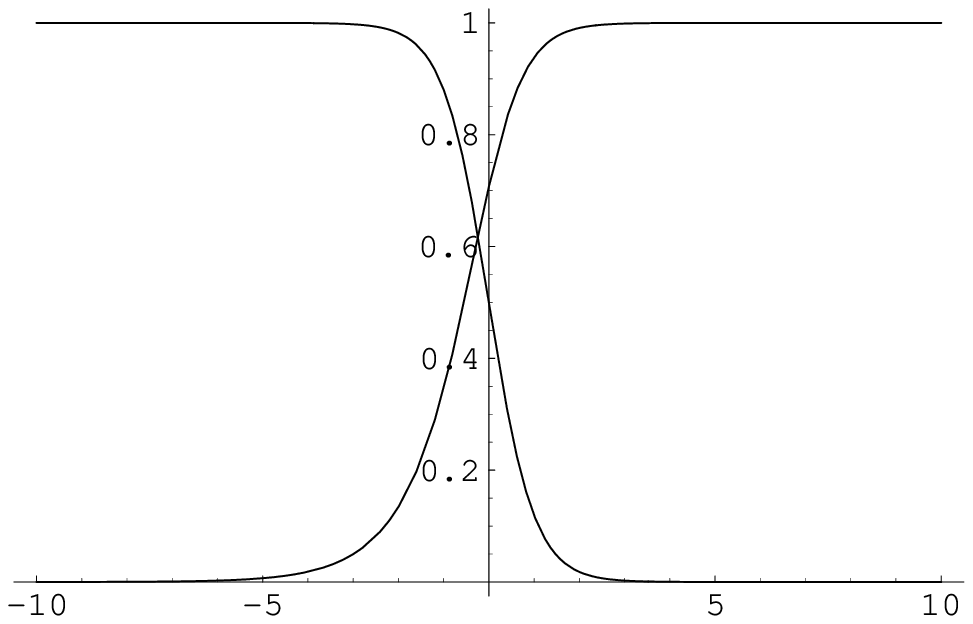}
\put(-10,10){$m_\h y$}
\put(-40,100){$g_e\phi (y)$ }\put(-70,40){$\lambda =1$}
\put(-170,100){$M(y)/m_\h$ }
\end{picture}
\caption{
The mass function $M(y)$ in 
Eq.(\ref{Mconfig2}) and 
the mode function $\phi(y)$ in 
Eq.(\ref{eq:massless-mode}). 
}
\label{fig1a}
\end{center}
\end{figure}

\begin{figure}[htb]
\begin{center}
\leavevmode
\epsfxsize=6cm
\epsfysize=4cm
\begin{picture}(100,100)(50,10)
\epsfbox{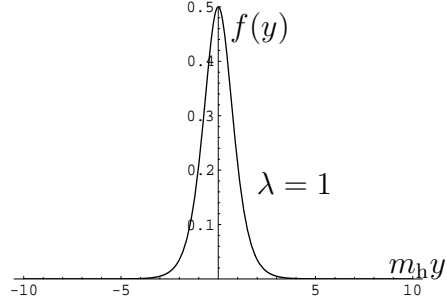}
\put(-20,10){$m_\h y$}
\put(-80,100){$f(y)$ }\put(-70,40){$\lambda =1$}
\end{picture}
\caption{
The profile function $f(y)$ of the energy density 
in Eq.(\ref{eq:exact-sol-profile}). 
}
\label{fig1b}
\end{center}
\end{figure}

To illustrate this localization mechanism by 
an explicit solution of our wall, 
we can take the exact solution (\ref{Mconfig}) as an example. 
Since we must satisfy the condition of 
vanishing 
$ M(y)
=\langle m_\t-g_\t \Sigma(y) \rangle $ 
as $y\rightarrow\infty$ (\ref{eq:tunig}), 
we require 
\begin{equation}
0=m_{\rm t}-g_{\rm t}m_{\rm h}/g_{\rm h}. 
\label{eq:tunig-h-v}
\end{equation} 
We will use $\lambda$ defined by 
$m_\t/m_\h=g_\t/g_\h\equiv \lambda /2>0$ 
together with $m_{\rm h}$ as the two 
independent parameters. 
We obtain 
\begin{eqnarray}
 M(y)&\!\!\!\equiv &\!\!\!m_\t-g_\t\langle \M\rangle 
=\myfrac{\lambda m_\h}2(1-\tanh(m_\h y))
={\lambda m_\h e^{-m_\h y}\over 2\cosh(m_\h y)} .
\label{Mconfig2}
\end{eqnarray} 
The mode function $\phi (y)$ and 
the energy density profile 
$f(y)$ are given by 
\begin{eqnarray}
\phi (y)
&\!\!\!=&\!\!\!
g_e^{-1}
\left({e^{m_\h y}\over 2\cosh(m_\h y)}\right)^{\frac{\lambda }2},
\label{eq:massless-mode}
\\
f(y)
&\!\!\!=&\!\!\!
{\lambda m_\h e^{(\lambda -1)m_\h y}\over 
2^\lambda \cosh^{(1+\lambda )}(m_\h y)},
\quad 
\left\{\begin{array}{cccc}
 f(y)&\rightarrow &{\lambda m_\h\over 2}\,e^{-2m_\h y} 
&\quad (y\rightarrow \infty )\\
f(y)&\rightarrow &{\lambda m_\h \over 2}\,e^{2\lambda m_\h y} 
&\quad (y\rightarrow -\infty )
\end{array}\right. .
\label{eq:exact-sol-profile}
\end{eqnarray}
The profile of these functions in 
Eqs.(\ref{eq:massless-mode}) and (\ref{eq:exact-sol-profile}) 
are illustrated in Fig.\ref{fig1a} and Fig.\ref{fig1b}, 
respectively. 

To close this subsection, we have to verify that our solution satisfies 
the boundary condition (\ref{condboundary}), 
since the mode function $\phi (y)$ does not vanish on the boundary. 
we can easily confirm the validity, by noting 
\begin{eqnarray}
 \delta B_{\mu\nu }\Big|_{\rm bondary}=2\partial _{[\mu}\delta A_{\nu]},
\end{eqnarray}      
and by performing a four-dimensional partial integration.

\subsection{Mass Spectrum and Four-Dimensional Effective Lagrangian}
\label{sc:massive-mode}

To obtain not only low-energy effective Lagrangian, but also 
the entire action on the background of our wall solution, it is 
necessary to work out all the massive modes. 
The result may also be of use in future study of the system. 
The basis vectors in momentum space of the massive states 
are three polarization vectors $\epsilon_\mu^i$ with $i=1, \cdots, 3$ 
defined by $p^\mu \epsilon_\mu^i=0$ together with the momentum $p^\mu$ itself. 
We substitute momentum-expansions of the fields $B_{M  N  }$ to 
the equation (\ref{eomB}) with the assumption $p^2\not=0$ to obtain 
\begin{eqnarray}
 B_{\mu\nu }(x,y)|_{\rm massive}
&\!\!\!=&\!\!\!\sum_{n\geq 1}\int {d^4p\over (2\pi i)^4}
\delta (p^2-m^2_{(n)})e^{ip_\lambda x^\lambda}
\left\{2ip_{[\mu}\epsilon _{\nu]}^ic_i(p){u'_{(n)}(y)\over M(y)}
+\epsilon ^i_\mu\epsilon _n^jd_{ij}(p)u_{(n)}(y)\right\}
, 
\label{eq:Bmn-massive}
\nn \\
 B_{\mu y}(x,y)|_{\rm massive}
&\!\!\!=&\!\!\!\sum_{n\geq 1}\int {d^4p\over (2\pi i)^4}
\delta (p^2-m^2_{(n)})m^2_{(n)}e^{ip_\lambda x^\lambda}
\epsilon _\mu^ic_i(p)\,{u_{(n)}(y)\over M(y)},
\label{eq:Bmy-massive}
\end{eqnarray}
where the function $u_{(n)}(y)$ is an eigenfunction of the following 
eigenvalue equation with an eigenvalue $m_{(n)}$, 
which serves as mass squared of the associated four-dimensional field 
\begin{eqnarray}
K u_{(n)}&\!\!\!=&\!\!\!m_{(n)}^2u_{(n)},\quad n=1,2,\cdots \nn
K&\!\!\!\equiv &\!\!\!-M(y){d\over dy}{1\over M(y)}{d\over dy}+M(y)^2.
\label{MassiveEq}
\end{eqnarray}
Note that the function $\phi (y)$ for the 
massless mode can be identified 
as the eigenfunction for zero eigenvalue : $K\phi =0$.  

Let us also consider modes of the scalars $\sigma ^\alpha $ 
of the tensor multiplets. 
The linearised equation of motion for the scalars 
$\sigma ^\alpha $ can be
read from the Lagrangian (\ref{eq:Taction2})  
\begin{eqnarray}
0
&\!\!\!=&\!\!\!M(y) \partial ^\mu\partial _\mu 
\sigma ^\alpha -\left(M(y){\sigma ^\alpha}'\right)'
-M(y)''\sigma ^\alpha 
+M(y)^3\sigma ^\alpha +{\left(M(y)'\right)^2\over M(y)}\sigma ^\alpha .
\end{eqnarray}
A similar argument using energy density shows that there is no 
massless modes in $\sigma^\alpha$. 
For massive modes, 
this equation can also be solved by using the function
$u_{(n)}(y)$ 
\begin{eqnarray}
 \sigma ^\alpha (x,y)
=\sum_{n}\int {d^4p\over (2\pi i)^4}
\delta (p^2-m^2_{(n)})e^{ip_\lambda x^\lambda}
\tilde \sigma ^\alpha (p)
{u_{(n)}(y)\over M(y)}.
\end{eqnarray}
This fact may be a result of the unbroken $D=4, {\cal N}=1$ SUSY.

To obtain a mass spectrum of the system, 
let us discuss the normalization and boundary conditions 
for massive state.  
In calculating the energy density of the 
system in  four dimensions, we encounter the following 
quantities,   
\begin{eqnarray}
 (u_{(n)},\,u_{(m)}),\quad (\phi ,\,u_{(n)}),\quad (u_{(n)},\,Ku_{(m)}), 
\cdots,
\end{eqnarray}
where the inner product $(u,\,v)$ is defined by,
\begin{eqnarray}
 (u,\,v)\equiv \int ^{\pi L}_{-\pi L}dy{u(y) v(y)\over M(y)}.
\label{eq:inner-product}
\end{eqnarray}
Since we eventually need to treat continuum of states, 
we will assume for reguralization purposes 
a compact space for the extra dimension $y$, 
as $-\pi L\leq y\leq \pi L$, and we 
will take the limit of $L\rightarrow \infty $ in the final stage.
Therefore, the normalization of the function 
$u_{(n)}$ must be defined 
by means of the inner product $(u_{(n)},\,u_{(m)})$. 
This inner product involving the operator $K$ 
in Eq.(\ref{MassiveEq}) has the following property 
\begin{eqnarray}
 (u,Kv)=\Delta (u,v)+(Ku,v), 
\qquad \Delta [u,v]\equiv 
\left[-{u v'\over M}+{u' v\over M}\right]^{\pi L}_{-\pi L}.
\end{eqnarray}
To make the operator $K$ hermitian $(u,Kv)=(Ku,v)$ 
with respect to the inner product (\ref{eq:inner-product}), 
we should demand that 
the contribution from the boundary, 
$\Delta [u,v]$, must vanish   
\begin{eqnarray}
\Delta [u_{(n)},u_{(m)}]
&\!\!\!=&\!\!\!\left[-{u_{(n)}u_{(m)}'\over M}
+{u_{(n)}'u_{(m)}\over M}\right]^{\pi L}_{-\pi L}=0,
\label{eq:nm-hermitian}
\\
\Delta [\phi ,u_{(n)}]
&\!\!\!=&\!\!\!\left[-\phi \left({u_{(n)}'\over M}
-u_{(n)}\right)\right]^{\pi L}_{-\pi L}=0 .
\label{eq:0n-hermitian}
\end{eqnarray}
Eq.(\ref{eq:0n-hermitian}) means that 
the massless mode 
is orthogonal to the massive modes. 
This is satisfied if the boundary 
conditions of the eigenfunction $u_{(n)}$ are given by 
\begin{eqnarray}
 {\cal B}_{(n)}(\pi L)={\cal B}_{(n)}(-\pi L)=0,
 \quad {\cal B}_{(n)}(y)\equiv 
u_{(n)}'(y)-M(y)u_{(n)}(y) . \label{Eq:BC} 
\end{eqnarray}
These boundary conditions are also enough to 
satisfy Eq.(\ref{eq:nm-hermitian}). 
With these conditions, the inner products of the 
eigenfunctions can be normalised as 
\begin{eqnarray}
 (u_{(n)},u_{(m)})=\delta _{n\,m},
 \quad (\phi ,u_{(n)})=0. 
\end{eqnarray}
On the other hand, 
the normalization of 
the massless mode function $\phi (y)$ is performed by 
\begin{eqnarray}
\int ^{\pi L}_{-\pi L}dyM\phi ^2
=\myfrac12\left[\phi ^2\right]^{\pi L}_{-\pi L}
\equiv {1\over 2g_e^2}, 
\label{phi-normalisation}
\end{eqnarray}
which we found in Sect.\ref{sc:massless-mode} as the 
contribution from the massless mode to the energy density. 
In fact, we find a divergent result 
 $(\phi ,\phi )=\infty $, if we apply the inner product 
 (\ref{eq:inner-product}) blindly also to the massless mode. 
We never encounter this quantity in the calculating 
the energy density. 
 With these normalization and the 
boundary conditions, we can compute the mass spectrum 
by a numerical analysis once the quantity 
$M(y)=\langle m_{\rm t}-g_{\rm t}\Sigma(y)\rangle $ 
is given. 
In the case of our exact solution (\ref{Mconfig2}) 
with $\lambda =1$, 
we can solve the equation (\ref{MassiveEq}) exactly 
and find the exact mass spectrum 
\begin{eqnarray}
 m_{(n)}=\sqrt{m_\h^2+\left({n\over 2L}\right)^2},\qquad n=1,2,\cdots ,
\end{eqnarray}    
which will be derived in the Appendix. 
In this case, we can explicitly see that 
the massless mode is always isolated from the massive mode 
even in the limit of $L\rightarrow \infty $, because of the mass gap. 
We expect that this desirable property will persist 
with other values of couplings and other configurations of the $M(y)$.  

Let us also describe the four-dimensional 
effective Lagrangian of the system to 
the second order of the fluctuations. 
Assuming that above mode functions $\phi (y), u_{(n)}(y)$ 
form a complete set to expand a function of $y$, we obtain 
expansions of the tensor fields $B_{M  N  }$ 
and the scalar fields $\sigma ^\alpha $ as  
\begin{eqnarray}
 B_{\mu\nu}(x,y)&\!\!\!=&\!\!\!\phi (y)F_{\mu\nu}(A(x))
+\sum_{n\geq 1}\left({u'_{(n)}(y)\over m_{(n)} M(y)}
F_{\mu\nu}(A^{(n)}(x))+u_{(n)}(y)C_{\mu\nu}^{(n)}(x)\right),
\\
 B_{\mu y}(x,y)
&\!\!\!=&\!\!\!
\sum_{n\geq 1}m_{(n)}{u_{(n)}(y)\over M(y)}A^{(n)}_\mu(x),
\\
\sigma ^\alpha (x,y)
&\!\!\!=&\!\!\!
\sum_{n\geq 1}{u_{(n)}(y)\over M(y)}\tilde \sigma ^\alpha _{(n)}(x).
\end{eqnarray}
Substituting this expansion to the Lagrangian, 
the quadratic terms of the four-dimensional effective Lagrangian can be read as
\begin{eqnarray}
 {\cal L}^{\rm eff}&\!\!\!=&\!\!\!{\cal L}_{\rm boundary}
+\sum_{n\geq 1}{\cal L}^{(n)}_{\rm massive}+{\cal L}_{\rm interactions}.
\end{eqnarray}
The first term is a contribution from the boundaries 
\begin{eqnarray}
 {\cal L}_{\rm boundary}
&\!\!\!=&\!\!\!-\myfrac14\left[\left\{\phi F_{\mu\nu}(A)
+\sum_{n\geq 1}
\left({u_{(n)}\over m_{(n)} }F_{\mu\nu}(A^{(n)})
+u_{(n)}C_{\mu\nu}^{(n)}\right)\right\}^2\right]^{\pi L}_{-\pi L}, 
\end{eqnarray}
where we used the boundary conditions. 
The second term corresponds to the kinetic term for massive states 
\begin{eqnarray}
 {\cal L}^{(n)}_{\rm massive}&\!\!\!=&\!\!\!
-{1 \over 4}F_{\mu\nu}(A^{(n)})^2
+\frac{m_{(n)}^2}2(A_\mu^{(n)})^2
+\frac1{12}F_{\mu\nu\lambda}(C^{(n)})^2
-\frac{m_{(n)}^2}4(C_{\mu\nu}^{(n)})^2
\nn
&\!\!\!&\!\!\!{}
+{1 \over 2}\partial ^\mu\tilde 
\sigma _{(n)}^\alpha \partial _\mu\tilde \sigma _{(n)}^\alpha 
-{m_{(n)}^2 \over 2}(\tilde \sigma _{(n)}^\alpha )^2.
\end{eqnarray}
The ${\cal L}_{\rm interactions}$ contains interactions which 
we do not consider here. 
If we take a limit of $L\rightarrow \infty $, 
we find that the term ${\cal L}_{\rm boundary}$
reduces to the kinetic term of the massless vector fields 
\begin{eqnarray}
 {\cal L}_{\rm boundary}\quad \rightarrow 
\quad {\cal L}_{\rm massless}=-\frac1{4g_e^2}F_{\mu\nu}(A)^2,
\end{eqnarray}
where, we used $u_{(n)}(y)\sim 1/\sqrt{L}$ at $y =\pm \pi L$ 
and thus, 
the values of $u_{(n)}$ on the 
boundaries vanish in this limit. 

\section{Four-Dimensional Coulomb Law 
}
\label{sc:coulomb}

As we explained in Sect.\ref{sc:massless-vector}, 
we demonstrated that the four-dimensional massless 
vector field $A_\mu$ generated from the solution of the tensor field 
$B_{M  N  }$ is localized on the BPS domain wall in this system. 
It is interesting and important to identify particles carrying the
 charge for the gauge transformation 
$\delta A_\mu=-\partial _\mu\Lambda $ of 
this massless localized vector 
field, which is not available in our present system. 
We expect that the gauge field $A_\mu(x)$ 
may be similar to the electromagnetic 
dual of the fundamental vector field $W_M  $ in our model. 
The fundamental vector field $W_M$ 
is the gauge field of the other $U(1)$ gauge 
transformation (\ref{Gtrf}) broken by the wall solution. 
Thjerefore we shall call the source for the fundamental 
vector 
field $W_M$ as ``electric'' and the source associated to the 
tensor field $B_{MN}$ as magnetic. 
In Ref.\cite{ShifmanYung}, the model is embedded into an 
${\cal N}=2$ SUSY $SU(2)$ gauge theory in four dimensions 
from the beginning. 
Therefore they were able to identify the source of the magnetic 
charge by incorporating classical solutions such as the 
Abrikosov-Nielsen-Olesen magnetic flux tube. 
In this way, they were able to show that the flux carried by 
the Abrikosov-Nielsen-Olesen magnetic flux tube 
becomes the source of the massless localized vector field 
that they found. 
In a similar spirit, it is an interesting and 
challenging task to generalize 
our model to non-Abelian gauge group such that the source 
of our massless localized vector field may be constructed 
as a classical solution, such as monopoles. 
Since we have not yet succeeded in building such a generalization, 
we will show only that our massless localized vector field 
does exhibit the four-dimensional Coulomb law between 
sources which we introduce here as external sources. 

Let us introduce a source $\JT^{M N }(x, y)$ 
for the tensor field $B_{MN}$ 
\begin{eqnarray}
{\cal L}
&\!\!\!=&\!\!\!{1\over 12M}F_{M  N L  }(B)F^{M  N L  }(B)
-{M\over 4} B_{M  N  }B^{M  N  }
+\myfrac12B_{M  N  }\JT^{M  N  }.
\end{eqnarray}
The equation of motion for the tensor field now reads 
\begin{eqnarray}
\partial ^L \left({1\over M}F_{M  N L  }(B)\right)
+MB_{M  N  }
=\JT_{M  N  }.\label{eq:EOMTsource}
\end{eqnarray}
By taking divergence, we obtain a source corresponding to the 
magnetic charge current $\tilde J^M(x, y) $ 
which is now introduced as an external source 
\cite{DubovskyRubakov}
\begin{eqnarray}
 \partial _N  (MB^{M  N  })=\partial _N  \JT^{M  N  }\equiv 
-\tilde J^M  ,
\quad \partial _M  \tilde J^M  =0. 
\label{eq:magnetic-current}
\end{eqnarray}

If we introduce the magnetic source $\tilde J^M(x, y) $ 
on the wall near $y=0$, but not in the bulk, 
it is reasonable to assume that the background is not 
disturbed by the source, and that only the massless mode 
is excited as in Eq.(\ref{eq:massless-comp})
\begin{eqnarray}
 B_{\mu\nu}(x, y)
 &\!\!\!=&\!\!\!\phi(y) F_{\mu\nu}(A(x)),
 \quad B_{\mu y}(x, y)=0.
\end{eqnarray}
The corresponding source $\JT_{MN}$ for the tensor field 
can be read from the equation of motion (\ref{eq:EOMTsource}) 
\begin{eqnarray}
 \JT_{\mu\nu}(x, y)&\!\!\!=&\!\!\!
0,\quad \JT_{\mu y}(x, y)
=\phi(y) \partial^\nu F_{\nu\mu}(A(x)).
\end{eqnarray}
Then Eq.(\ref{eq:magnetic-current}) implies 
the following distribution of the magnetic charge current 
$\tilde J_M(x, y)$  
\begin{eqnarray}
 \tilde J_{\mu}(x, y)
=\phi'(y) \partial^\nu F_{\nu\mu}(A), 
\quad 
 \tilde J_{4}(x, y)=\partial^\nu \JT_{\nu y}=0 
.
\end{eqnarray}
Now we can view the first equation as the usual equation for 
the source exciting our massless gauge field 
\begin{eqnarray}
\partial ^\nu F_{\mu\nu}(A(x))
=
-g_e^2J_\mu(x), 
\label{eq:maxwell-source}
\end{eqnarray}
where the source current for our gauge field 
$J_\mu(x)$ is defined in terms of the magnetic 
source current 
$\tilde J_M(x, y)$ as 
\begin{eqnarray}
 \tilde J_\mu(x, y)
 =g_e^2\phi'(y) J_\mu(x),
 \quad \tilde J_4(x, y)=0.
\end{eqnarray}
We can see that this configuration is consistent 
with the precondition of putting the 
magnetic source on the wall, since the function 
$\phi'(y)=M(y)\phi(y)$ is localized on the wall.   
If we take 
 a static point source for the massless localized vector field 
 $J_\mu(x)=\delta _\mu^0\delta ^3(x)$, 
 we can easily see from the above Eq.(\ref{eq:maxwell-source}) 
that four-dimensional 
Coulomb law of the usual minimal electromagnetic interaction 
is reproduced. 

On the other hand, 
the source $\JT_{MN}(x, y)$ for the tensor multiplet 
which causes this configuration is given by 
\begin{eqnarray}
 \JT_{\mu\nu}(x, y)&\!\!\!=&\!\!\!
0,\quad \JT_{\mu y}(x, y)=g_e^2\phi(y) J_\mu(x).
\end{eqnarray}
It is interesting to note 
the field $\phi(y)$ approaches a nonvanishing 
constant value asymptotically as illustrated in 
Fig.\ref{fig1a}. 
This behavior of the tensor field source 
$\JT_{\mu y}(x, y)$ appears to 
suggest a certain flux coming out of 
the brane to positive infinity $y= \infty $. 
However, we believe that 
this should be a  fictitious flux like a Dirac string 
for a monopole, since the energy density corresponding 
to the massless vector excitation is localized around the wall, 
as shown in Eq.(\ref{eq:energy-density-massless}). 
If we wish to place the magnetic source 
$\tilde J^M(x, y) $ in the bulk, 
we need to take into account of the deformation 
of the background 
due to the presence of the magnetic source \cite{DubovskyRubakov}. 
We wish to investigate the nature of the massless vector field and 
its coupling further in subsequent publications.

\section{Discussion}
\label{sc:discussion}

We can extend our mechanism for a massless localized gauge field 
on a wall to an arbitrary space-time dimensions, 
provided we ignore SUSY for the moment. 
Suppose that we have the same bosonic Lagrangian as our 
Lagrangian $\left.{\cal L}_{\rm wall}\right|_{\rm bosonic}$ 
in (\ref{eq:wall-lagrangian}) to build a wall in arbitrary $D$ 
space-time dimensions. 
Let us add the following Lagrangian for a $(D-3)$-form field 
$B_{\mu _1 \cdots \mu _{D-3}}$ 
instead of Eq.(\ref{eq:quad-tensor-lag}) 
in five dimensions 
\begin{eqnarray}
 {\cal L}_{(D-3){\rm form}}
&\!\!\!=&\!\!\!{1\over 2(D-2)! M}
F_{\mu _1 \cdots \mu_{D-2}}(B)
F^{\mu _1 \cdots \mu _{D-2} }(B)
-{M \over 2(D-3)!} 
B_{\mu _1 \cdots \mu _{D-3} }B^{\mu _1 \cdots \mu _{D-3}}, 
\label{eq:D-2formLag}
\end{eqnarray}
\begin{equation}
 {\cal L}_{\rm total}=\left.{\cal L}_{\rm wall}\right|_{\rm bosonic}
+{\cal L}_{ (D-3){\rm form}} . 
\label{eq:D+1ToTaction}
\end{equation}
We expect that the same mechanism may be operative in this system as well: 
namely a massless localized $D-4$ form field $A_{\mu _1 \cdots \mu _{D-4}}$ 
is likely to be contained in the $D-3$ form field 
$B_{\mu _1 \cdots \mu _{D-3}}$ as 
$B_{\mu _1 \cdots \mu _{D-3}}(x, y)
=\phi(y)(D-3)\partial_{[\mu _1}A_{\mu _2 \cdots \mu _{D-3}]}(x)$, 
where $\phi(y)$ is a mode function of the massless form field. 
However, we should note that it may or may not be realized with SUSY, 
since the constraint of SUSY in higher dimensions are quite severe. 
We can think of the above Lagrangian just a bosonic model without SUSY, 
although it is motivated from SUSY models. 
The electromagnetic dual field of the fundamental vector 
field $W_M $ in 
$D$ dimensions should be a $D-3$ form, and the 
electromagnetic dual of $W_M $ 
in $D-1$ dimensions should be a $D-4$ form. 
Therefore the fundamental $D-3$ form field $B$ 
and its massless localized component of  $D-4$ form 
field $A$ precisely possesses the expected degree of forms. 

Let us finally list some of open problems 
for future research. 

It is most desirable to be able to obtain charged fields 
which interact with our massless localized gauge field. 
This may be achieved by introducing a non-Abelian generalization 
of our model. 
Therefore it is an interesting open problem to make our model 
non-Abelian, such as $SU(2)$. 
This might answer the question whether our massless localized 
gauge field is really an electromagnetic 
dual of the fundamental gauge field. 
It may hopefully lead to more realistic model building 
with the ${\cal N}=1$ SUSY standard model matter content 
\cite{DGSW}. 

It would be interesting to understand more deeply the symmetry 
or topological reason for the existence of massless localized 
gauge field. 

We had to make one fine-tuning among parameters of the 
hypermultiplet and 
tensor multiplet (\ref{eq:tunig}) or (\ref{eq:tunig-h-v}) 
to obtain a massless localized gauge field. 
It is also an interesting open question to understand or explain 
this fine-tuning from other argument. 
If this question can be addressed successfully, it may also be 
possible to fix other parameters of our model, such as 
$\lambda/2\equiv g_t/g_\h=m_\t/m_\h$. 

It should be straight-forward to embed our system into supergravity 
in five dimensions \cite{KugoOhashi, EFNS, AFNS}. 
Then there are of course interesting questions to be explored, such as 
the fate of graviphoton. 

Although we have obtained the four-dimensional gauge 
coupling as a topological charge, it can still be 
compatible with the concept of running coupling 
due to quantum effects. 
We can draw an interesting analogy to the fact that 
the masses of the BPS 
magnetic monopole and dyon are characterized 
as topological charge, which appear as the central charge 
in the SUSY algebra. 
The exact solution of the 
${\cal N}=2$ SUSY gauge theories \cite{SeibergWitten} 
demonstrated explicitly that these topological 
charges receive interesting 
nonperturbative effects from quantum loops. 
It is a challenging future problem to consider 
quantum effects in our theory. 

\renewcommand{\thesubsection}{Acknowledgments}
\subsection{}

One of the authors (N.S.) acknowledges a useful 
discussion of gauge field localization and tensor multiplet 
with E.Kh.~Akhmedov, Bernard de Wit, Gia Dvali, Nobuhito Maru, 
Seif Randjbar-Daemi, and Roberto Soldati. 
He also thanks the hospitality of the International Centre for 
Theoretical Physics at the last stage of this work. 
This work is supported in part by Grant-in-Aid 
 for Scientific Research from the Japan Ministry 
 of Education, Science and Culture 13640269 and 01350. 
The work of K.O. is supported in part by Japan Society for the Promotion
of Science under the Post-doctoral Research Program.

\renewcommand{\thesubsection}{\thesection.\arabic{subsection}}

\appendix

\section{Massive modes}
\label{sc:Massive-modes}
Let us consider the solution of the equation of motion for the massive
mode (\ref{MassiveEq}) with the boundary condition (\ref{Eq:BC}).
A canonical mode functions 
$v_{(n)}(y)\equiv u_{(n)}(y)/\sqrt{M}$ may be more convenient 
than the original mode functions $u_{(n)}(y)$, 
because of the definition of the inner product (\ref{eq:inner-product}).    
Rewriting the equation of motion (\ref{MassiveEq}) by the canonical
mode functions $v_{(n)}(y)$, 
we obtain an ordinary Schr\"odinger equation 
\begin{eqnarray}
\left(-{d^2\over dy^2}+V(y)\right)v_{(n)}(y)=m_{(n)}^2v_{(n)}(y),
\end{eqnarray}
where the potential $V(y)$ is given by 
\begin{eqnarray}
 V(y)=M(y)^2-\sqrt{M(y)}\left({\left(\sqrt{M(y)}\right)'\over M(y)}\right)'.
\end{eqnarray}
If we use the configuration (\ref{Mconfig2}) 
for the quantity $M(y)$,
the $V(y)$ can be read as 
\begin{eqnarray}
 V(y)=
 {m^2_\h\over 4\cosh^2(m_\h y)}
 \left(2+(1+\lambda^2)\cosh(2m_\h y)
+(1-\lambda^2)\sinh(2m_\h y)\right).
\end{eqnarray}
The potential $V(y)$ as a  function of $y$ 
approaches the 
value $m_\h^2$ asymptotically 
at $y\rightarrow \infty$, and approaches the
value $\lambda^2 m_\h^2$ at the opposite infinity.
 Therefore, we find
that the mass gap between the massless mode and the massive mode is 
$\lambda m_\h$ for the case $\lambda\leq 1$, whereas the mass gap is
$m_\h$ for the case $\lambda>1$. 

\begin{figure}[htb]
\begin{center}
\leavevmode
\epsfxsize=8cm
\epsfysize=4cm
\begin{picture}(100,110)(50,10)
\epsfbox{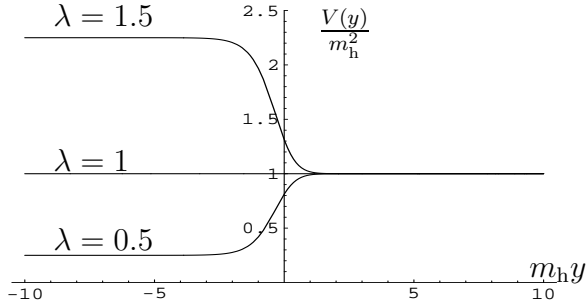}
\put(-20,10){$m_\h y$}
\put(-100,100){${V(y)\over m_\h^2}$ }\put(-200,50){$\lambda =1$}
\put(-200,105){$\lambda =1.5$}\put(-200,20){$\lambda =0.5$}
\end{picture}
\caption{The potential $V(y)$ as a function of $y$ 
for $\lambda=0.5, 1, 1.5$. 
}
\label{fig2}
\end{center}
\end{figure} 
We note that the potential becomes 
a constant in the case of $\lambda=1$ : 
 $V(y)=m_\h^2$. 
 Therefore we can fortunately solve the equation of motion 
 exactly and obtain the mass spectrum for $\lambda=1$. 
In this case, the mass function 
$M(y)=\langle m_{\rm t}-g_{\rm t} \Sigma(y) \rangle$ 
is given by the configuration of the vector multiplet 
scalar $\Sigma$ in Eq.(\ref{Mconfig}) as 
\begin{eqnarray}
M(y)={m_\h\over 2}\left\{1-\tanh(m _\h(y-y_0))\right\} >0,
\end{eqnarray}
where we have restored an arbitrary parameter $y_0$ 
corresponding to the position of the wall. 
Exact solutions of the equation of motion (\ref{MassiveEq})
are given by 
\begin{eqnarray}
u_{(0)}(y)&\!\!\!\equiv &\!\!\!
\phi (y)={e^{\frac{m_\h}2 (y-y_0)}\over g_e(L)\sqrt{2\cosh(m_\h(y-y_0))}}
=\sqrt{M(y)\over m_\h g_e^2(L)}\,e^{m_\h(y-y_0)},\quad  m_0=0,\nn
u_{(n)}(y)&\!\!\!=&\!\!\!
C_{(n)}\sqrt{M(y)}\cos(\tilde m_{(n)} (y-y_{(n)})), 
\quad m_{(n)}=\sqrt{m_\h^2+\tilde m_{(n)}^2},\quad n=1,2,\cdots , 
\end{eqnarray}
where a parameter $g_e(L)$ is defined by the normalization 
(\ref{phi-normalisation}) and reduces to 
the effective four-dimensional gauge coupling 
$g_e$ in the limit $L\rightarrow \infty$, and
$y_{(n)},\,\tilde m_{(n)} $ are arbitrary constant parameters.
In this case, the quantity ${\cal B}_{(n)}(y)$ defined by 
Eq.(\ref{Eq:BC}) is given by 
\begin{eqnarray}
{\cal B}_{(n)}(y)
&\!\!\!=&\!\!\!-C_{(n)}\sqrt{M(y)}
\left(m_\h \cos(\tilde m_{(n)}(y-y_{(n)}))
+\tilde m_{(n)} \sin(\tilde m_{(n)}(y-y_{(n)}))\right).
\end{eqnarray}
Thus, the boundary conditions (\ref{Eq:BC}) become 
\begin{eqnarray}
 \tan(\tilde m_{(n)}(\pi L-y_{(n)}))=-{m_\h\over \tilde m_{(n)}},
\qquad 
\tan(\tilde m_{(n)}(\pi L+y_{(n)}))={m_\h\over \tilde m_{(n)}},
\end{eqnarray}
which determine the parameters $y_{(n)},\, \tilde m_{(n)}$ as 
\begin{eqnarray}
\tilde m_{(n)}={n\over 2L},\qquad 
 y_{(n)}=
L\left({2\over n}\arctan\left({2m_\h L\over n}\right)-\pi \right),\qquad 
n=1,2,\cdots .
\end{eqnarray}
Note that a massive mode corresponding to $n=0$ is prohibited by 
the boundary condition, whereas the massless mode is permitted.
Therefore the mass spectrum can be read as
\begin{eqnarray}
 m_{(n)}=\sqrt{m^2_\h+\left({n\over 2L}\right)^2},\qquad n=1,2,\cdots
\end{eqnarray}
The normalization constants are found to be 
\begin{eqnarray}
1&\!\!\!=&\!\!\!\int ^{\pi L}_{-\pi L}dy{u_{(n)}(y)^2\over M(y)}=
C_{(n)}^2\int ^{\pi L}_{-\pi L}dy'\cos^2\left({n\, y'\over 2 L}\right)
=\pi LC_{(n)}^2.
\end{eqnarray}     
Thus we obtain the exact solutions for the massive modes 
\begin{eqnarray}
 u_{2n}(y)&\!\!\!=&\!\!\!
\sqrt{M(y)\over \pi L}\cos\left({n\, y\over L}
-\arctan\left({m_\h L\over n}\right)\right),\qquad n=1,2,3,\cdots\nn
 u_{2n+1}(y)&\!\!\!=&\!\!\!
\sqrt{M(y)\over \pi L}\sin\left({(2n+1) y\over 2L}
-\arctan\left({2 m_\h L\over 2 n+1}\right)\right),\qquad n=1,2,3,\cdots.
\end{eqnarray}

\newcommand{\J}[4]{{\sl #1} {\bf #2} (#3) #4}
\newcommand{\andJ}[3]{{\bf #1} (#2) #3}
\newcommand{\AP}{Ann.\ Phys.\ (N.Y.)}
\newcommand{\MPL}{Mod.\ Phys.\ Lett.}
\newcommand{\NP}{Nucl.\ Phys.}
\newcommand{\PL}{Phys.\ Lett.}
\newcommand{\PR}{ Phys.\ Rev.}
\newcommand{\PRL}{Phys.\ Rev.\ Lett.}
\newcommand{\PTP}{Prog.\ Theor.\ Phys.}
\newcommand{\hep}[1]{{\tt hep-th/{#1}}}

\end{document}